\documentclass[prb,aps,twocolumn ,10pt,superscriptaddress,showpacs]{revtex4}
\usepackage{graphicx}
\usepackage[caption=false]{subfig}
\usepackage{bm}
\usepackage{epsfig}
\usepackage{amsmath}
\usepackage{amssymb} 
\usepackage{comment}
\usepackage{hyperref}
\usepackage[usenames,dvipsnames]{color}
\hypersetup{urlcolor=blue, colorlinks=true, citecolor=blue, linkcolor=blue}

\newcommand{\bpm}{\begin{pmatrix}}
\newcommand{\epm}{\end{pmatrix}}
\newcommand{\ba}{\begin{eqnarray}}
\newcommand{\ea}{\end{eqnarray}}

\newcommand{\mysection}[1]{\section{ #1}}

\begin{document}

\title{Fractional boundary charges in quantum dot  arrays with density modulation}

\author{Jin-Hong Park}
\affiliation{RIKEN Center for Emergent Matter Science, Wako, Saitama 351-0198, Japan}

\author{Guang Yang}
\affiliation{RIKEN Center for Emergent Matter Science, Wako, Saitama 351-0198, Japan}

\author{Jelena Klinovaja}
\affiliation{Department of Physics, University of Basel, Klingelbergstrasse 82, CH-4056 Basel, Switzerland}

\author{Peter Stano}
\affiliation{RIKEN Center for Emergent Matter Science, Wako, Saitama 351-0198, Japan}
\affiliation{Institute of Physics, Slovak Academy of Sciences, 845 11 Bratislava, Slovakia}

\author{Daniel Loss}
\affiliation{RIKEN Center for Emergent Matter Science, Wako, Saitama 351-0198, Japan}
\affiliation{Department of Physics, University of Basel, Klingelbergstrasse 82, CH-4056 Basel, Switzerland}

\date{\today}

\begin{abstract}

We show that fractional charges can be realized at the boundaries of a linear array of tunnel coupled quantum dots in the presence of a periodically modulated onsite potential. While the charge fractionalization mechanism is similar to the one in polyacetylene, here the values of fractional charges can be tuned to arbitrary values by varying the phase of the onsite potential or the total number of dots in the array.  We also find that the fractional boundary charges, unlike the in-gap bound states, are stable against static random  disorder. We discuss the minimum array size where fractional boundary charges can be observed. 

\end{abstract}


\maketitle

\mysection{Introduction}
Charge fractionalization is a striking emergent phenomenon that can take place in correlated electronic systems. In two dimensions, quasiparticle excitations in fractional quantum Hall liquids carry fractional charges,\cite{laughlin82,depicciotto97,saminadayar97} which along with the quantization of magnetic flux leads to exotic fractional exchange statistics.\cite{Halperin84,Wilczek84} In one dimension, fractionalized charge excitations were observed in transport measurements \cite{steinberg07,kamata14,inoue14} in quantum wires and coupled edge channels of integer quantum Hall states. A third class constitutes 
states in one-dimensional dimerized polymers,
first considered by Su, Schrieffer, and Heeger (SSH).~\cite{ssh79} There, a soliton configuration of the lattice deformation produces a gap in the spectrum and binds a nondegenerate fermionic zero-mode, as discovered initially in 
a continuum model by Jackiw and Rebbi.\cite{jr76}
This zero-energy in-gap bound state is associated with a well-defined \cite{fnt1} half-integer charge, the quantization of which is 
protected by chiral symmetry.\cite{js81} 

Though the half-integer charge gives rise to unusual spin-charge relations of solitons in polyacetylene,\cite{ssh79} there its direct access is hindered by the spin degeneracy. As noticed early on, breaking the chiral symmetry offsets the soliton-mode energy from zero and its charge from 1/2.~\cite{ss81,gw81}
Extensions of the SSH model in this regard have been considered, relying on diatomic polymers \cite{rm82,js83} or multiple lattice modes,\cite{kivelson83} though these rather involved constructions remained without experimental realizations.

These ideas were recently reconsidered, aiming both at platforms with more advanced experimental controls \cite{spinhall,goldman2010:PRL, Kraus12,Gangadharaiah2012,poshakinskiy2014:PRL} and theoretical generalizations of the model by including interactions\cite{Gangadharaiah2012,budich2013:PRB,xu2013:PRL,zhu2013:PRL,grusdt2013:PRL} and to higher dimensions.\cite{hou07,swm08,rf11,Szumniak15} Even more interestingly, half-integer charges associated with in-gap bound states have been predicted in novel topological phases as precursors of exotic topological matter. \cite{KSLPRL2012,KLPRL2013,KLPRL2014,wakatsuki2014:PRB,rainis2014:PRL} 
However, the energy of these in-gap bound states, localized at the boundaries of the chain, is sensitive to disorder.\cite{Gangadharaiah2012,madsen2013:PRB,EP} 
Remarkably, we find here that this is not the case for the fractional boundary charge, which remains stable in the presence of disorder. 
This fractional boundary charge gets contributions from all occupied fermion states (affected by the boundary) which might or might not include  in-gap states depending on system parameters.
Qualitatively, this can be understood in terms of the stability of a band insulator, where charges  can be displaced by local fields only to a limited distance, while shifts in energy levels could be substantial.
A crucial ingredient in our model is the absence of the chiral symmetry of the SSH model,~\cite{ssh79} which would otherwise make the spatial profiles of the fractional boundary charges and the in-gap bound states identical.\cite{js81} 
The lack of this symmetry allows the two quantities to be independent and respond differently to disorder: the fractional boundary charges are robust while the in-gap bound states can be pushed all the way into the band continuum and completely delocalize. 

\begin{figure}
\includegraphics[width=85mm]{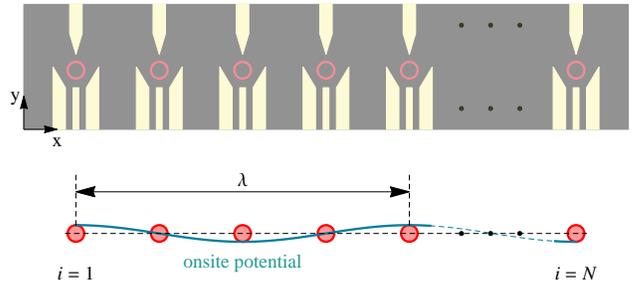}
\caption{(color online) 
Schematic view of a linear array of $N$ tunnel coupled quantum dots (red circles; indexed by $i$) under a periodically modulated onsite potential (blue curve) with period $\lambda$.
}
\label{fig:scheme}
\end{figure}

We demonstrate these discoveries in a tight-binding model, envisaged to be realized in an array of electrically tunable coupled quantum dots (QDs).
The recent progress in fabrication and control\cite{Hanson2007,Kloeffel2013}
 motivates us to consider the linear QD array as a realistic platform where such fractional boundary charges can be established and probed experimentally. A periodically modulated onsite potential induces fractional charges at the boundaries of the array, 
with the values controlled by the  phase $\theta$ of the potential and the number of QDs in the array. These two easily controllable parameters allow for, respectively, continuous and discrete variations of the fractional charge values.
We note that the same physics can also be realized in other platforms such as nanowires with superlattice structures. \cite{Gangadharaiah2012}

The outline of the paper is as follows. In Sec.~II, we introduce the model of a QD array. In Sec.~III, we discuss requirements for a consistent definition of fractional charges, and demonstrate (with details in App.~A) that these are fulfilled in our model. In Sec.~IV we present values of fractional boundary charges obtained from numerics showing that these are tunable by experimentally accessible parameters, and explain the observed results by analytical derivation. In Sec.~V, we show that the fractional boundary charges are stable against disorder. Finally, in Sec.~VI we estimate the minimal size of an array where fractional boundary charges could be observed experimentally.

\mysection{Model}
We consider a linear array of $N$ tunnel coupled QDs with a gate-induced periodic potential modulation, as illustrated in Fig.~\ref{fig:scheme}, and described by the Hamiltonian
\begin{equation} 
H = -t \sum_{i=1}^{N-1} (c^{\dag}_i c_{i+1}  + c^{\dag}_{i+1} c_i ) + \Delta \sum_{i=1}^N \cos ({2 \pi \over \lambda} i + \theta) c^{\dag}_i c_i, \label{eq:H}
\end{equation} 
with $c_i$ being the annihilation operator of an electron in the $i$th QD and $t$ the hopping amplitude. The potential modulation has strength $\Delta$, period $\lambda$, and phase offset $\theta$. We neglect the electron-electron, as well as spin-dependent interactions (such as spin-orbit and hyperfine effects) and omit spin indexes. We will reinstate the spin degree of freedom when necessary.

\begin{figure}
\includegraphics[width=85mm]{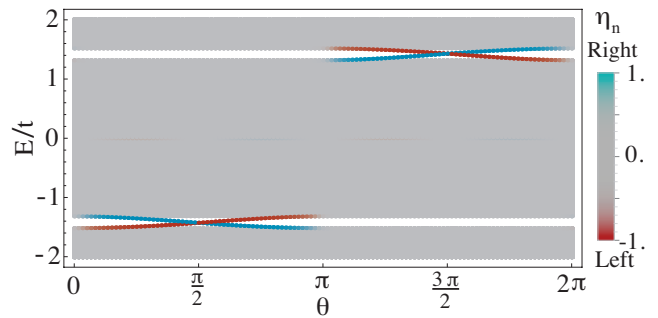}
\caption{(color online) 
Energy spectrum of $H$ in Eq.~(\ref{eq:H}) for vanishing boundary conditions,  $N = 401$, $\lambda = 4$, and $\Delta/t = 0.2$, with $\Delta$ the potential amplitude and $t$ the tunneling amplitude. For the state 
$n$ (wavefunction $\psi_n$),
the energy is plotted as a dot with color hue defined by normalized center of mass 
$\eta_n =  2\sum_i |\psi_n(i)|^2/N-1$. 
A state localized on the left (right) boundary has $\eta\approx -1 (+1)$, while an extended state has $\eta\approx 0$. 
} 
\label{fig:spectrum}
\end{figure}

The spectrum of $H$ is plotted in Fig.~\ref{fig:spectrum} for a representative choice $\lambda=4$. The potential modulation opens gaps, 
inside which the well known in-gap (Tamm~\cite{tamm} or Shockley~\cite{shockley}) states reside for a ($\lambda$ and gap dependent) range of values of $\theta$.~\cite{Gangadharaiah2012} In addition to being localized at the array edges, these states differ from the rest of the spectrum by a distinctive dispersion; upon changing $\theta$ they cross the gap. 

\mysection{Fractional charge definition}
In a many-body system of integer charges (in units of elementary charge $e$) the definition of a {\it fractional} charge necessarily requires to consider  differences between charges defined for different system configurations. 
To be more concrete, when we measure a local charge density of interest  
we subtract  a constant background, such as the bulk contribution, and then compare this local charge density for different configurations. 
\footnote{In linearized models the removed quantity is infinite,\cite{jr76} making the procedure less intuitive, but our model is free from such complications as the bulk density is finite.} 

The following operator describes such a charge measurement with subtracted background,
\begin{equation}
\hat {Q}_f=\sum_{i=1}^{N}f_i(e c_i^{\dagger}c_i-\bar\rho).
\label{eq:charge operator}
\end{equation}
Here, $\overline{\rho}$ is the bulk charge density and $f_i\in \langle 0,1\rangle$ is the profile function defining which part of the system is being measured by the operator $\hat{Q}_f$. For concreteness, the left end of the array can be associated with $\hat{Q}_L$ by taking
\begin{align}
 f^L_i = \left\{ \begin{array}{ll}
         1, & \mbox{if $i<l_0$},\\
        1- {i-l_0 \over W}, & \mbox{if $l_0 \leq i \leq l_0 + W$},\\
        0, & \mbox{if  $i > l_0 + W $}. \end{array} \right. 
        \label{eq:profile}
\end{align} 
Here, $l_0$ defines which parts of the system contribute to $\hat{Q}_L$, while  $W$ characterizes the cut-off, with larger $W$ meaning a smoother profile drop. On the other hand, the bulk density $\overline{\rho}$ is fixed by the chemical potential $\mu$. Setting it inside the lower gap gives $\overline{\rho}=e/\lambda$. This can be understood by considering periodic boundary conditions, where this choice means one occupied band, out of the total $\lambda$ bands in the Brillouin zone.~\footnote{The original Brillouin zone of free electrons $k\in \langle -\pi, \pi \rangle$ is shrunk $\lambda$ times to $k\in \langle -\pi/\lambda, \pi/\lambda \rangle$ by applying the modulation potential with period $\lambda$.} With these three parameters implicitly included, the fractional boundary charge is defined as the expectation value $Q_f=\langle \hat{Q}_f\rangle$ in the system ground state,~\footnote{We assume the temperature is much smaller than the gap $\Delta$, so that a zero temperature limit can be assumed for simplicity.}
\begin{equation}
Q_f = \sum_i f_i ( \rho_i - \overline{\rho} ).
\label{eq:charge}
\end{equation}
Here, $\rho_i=e\langle c^\dagger_i c_i\rangle$ is the ground state charge density,
while $f_i\equiv f_i^L$ gives the left boundary charge, and similarly $f_i\equiv f^R_i=f^L_{N-i}$ defines the right boundary charge.

A well-defined fractional charge requires two non-trivial properties. First, $Q_f$ must be independent of the details of the profile function $f$, meaning here both $l_0$ and $W$. For this the local density has to converge fast enough to its bulk value upon moving away from the wire end. 
As we will see below, in our model the convergence is exponential.
Second, the fractional (non-integer) value of $Q_f$ must not arise as an average over integral values. This can be cast as a condition on the standard deviation of the charge operator to be negligible compared to its mean, $ \delta Q_f \equiv \langle (\hat{Q}_f - Q_f)^2 \rangle^{1/2} \ll Q_f$. Similarly as for $Q_f$, the bulk charge contributes also to $\delta Q_f$, increasingly for a more abrupt profile drop. This contribution can be suppressed only in the limit $1/W \to 0$, as is well known.~\cite{sharpness} Without necessarily being experimentally accessible, this limit allows one to single out the intrinsic quantum fluctuations of the charge $Q_f$. We confirmed that in our model the charges defined by Eq.~\eqref{eq:charge operator} indeed correspond to well defined sharp quantum observables, fulfilling both requirements stated above. These consistency checks, analogous to previously investigated models, are presented in App.~\ref{app:check}.

\begin{figure}
\includegraphics[width=85mm]{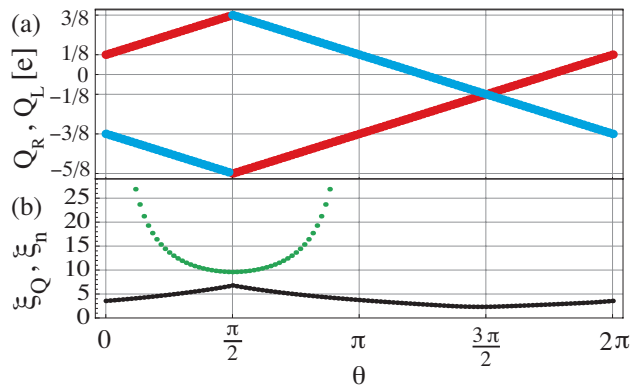}
\caption{(color online) (a) Left ($Q_L$, red) and right ($Q_R$, blue)  fractional boundary charges numerically obtained as function of $\theta$ for an array with $N = 401$ sites, $\lambda = 4$, $\Delta/t = 0.2$, $l_0=200$, $W=\lambda$, and the chemical potential set at the crossing of the in-gap states. 
(b) Inverse participation ratios of 
the lower in-gap state if it exists (green; $\xi_{n}^{-1}=\sum_{i=1}^N |\psi_i|^4$ with $\psi$ the in-gap bound state wavefunction) and
$Q_L$ (black; $\xi_{Q}^{-1}$ is given by an analogous formula with $|\psi_i|^2$ replaced by the averaged density difference, see Eq.~\eqref{eq:IPR} in App.~\ref{app:check}).
} \label{fig:charge-theta}
\end{figure}

\mysection{Fractional charge values}
We plot the left and right boundary charges defined by Eq.~\eqref{eq:charge} in Fig.~\ref{fig:charge-theta}(a) and obtain striking behavior: the boundary charges are fractional, depend linearly on the phase $\theta$, and do not show any direct relation to the in-gap states, the latter of which exist in the lower band gap only for $\theta \in\langle 0, \pi\rangle$ (see Fig.~\ref{fig:spectrum}). 
The independence of the two quantities is further corroborated by comparing the localization lengths of the boundary charge $\xi_Q$ and in-gap state $\xi_n$, plotted in Fig.~\ref{fig:charge-theta}(b). Not only do they differ from each other, they even evolve oppositely upon changing $\theta$: the in-gap bound states are maximally localized at $\theta=\pi/2$ where they cross, whereas the boundary charges are maximally extended at this $\theta$-value. In particular, there is no special feature visible in the value or localization behavior of the boundary charges at the points $\theta=0$ and $\theta=\pi$, where the in-gap states merge into the bulk and delocalize with $\xi_n\sim N$ [out of the range shown in Fig.~\ref{fig:charge-theta}(b)]. With additional differences demonstrated below, we come to our first important finding: unlike in the SSH model,~\cite{ssh79} where the boundary charges and in-gap states have identical spatial profiles, here these two quantities are totally different. 

We next turn to an analytic discussion of the boundary charges $Q_L$ and $Q_R$ and derive explicitly their linear $\theta$-dependence seen in the numerics.
To begin with, we split the total charge of the array as
\begin{equation}
Q_{tot} \equiv \sum_i \rho_i = Q_{bulk} + Q_L +Q_R,
\label{eq:moae}
\end{equation}
where we define the bulk charge as $Q_{bulk}=N \overline{\rho}$. We use here the fact that the average charge density per site ${\bar \rho}=e/\lambda$ 
for a free system does not change 
when the potential is turned on (the case under consideration),
however,  the charge distribution becomes non-uniform close to the boundary in the presence of the potential.
The chemical potential is in the gap such that  $Q_{tot}/e$ is an integer, equal to the number of occupied states. 
The boundary charges $Q_{L,R}$ are local quantities with localization lengths much smaller than the array length (see Fig.~\ref{fig:charge-theta}(b)).
For example, $Q_L$ ($Q_R$) depends on the potential shape only close to the left (right) end of the array.

First we imagine that we extend the array by one site on the {\it right} end, which means that $N\to N+1$ or equivalently that $\delta \to (\delta +1)\ {\rm mod}\, \lambda$, where we introduced the integer $\delta =(N+1)\ {\rm mod}\, \lambda$ to ease the notation.
This extension does not affect the charge on the opposite (left) end $Q_L$, 
which means that $Q_L$ can only be a function of $\theta$ but not of $\delta$, {\it i.e.}, $Q_L=Q_L(\theta)$. Furthermore, we note that adding one site increases the bulk charge by $e/\lambda$. To keep $Q_{tot}/e$ an integer, $Q_R$ must therefore decrease by the same amount 
\begin{equation}
\delta Q_R=-e/\lambda,
\label{delta_charge}
\end{equation}
while $Q_L$ remains unchanged as it is not influenced by changes on the right end. 

Next, we shift the phase $\theta$ such that the right end is identical to the situation  before the extension. Since the period of the cosine potential $\cos(2\pi i/\lambda +\theta)$ is given by $\lambda$, this means that we need the shift $\theta \to \theta -2\pi/\lambda$, where  $2\pi/\lambda$ is the phase change over one site. Under these two shifts, the charge on the right end $Q_R$ is invariant (since we are back to the same physical situation at the right end). 
On the other hand, $Q_R$ depends in general on $\delta$ and $\theta$. However, since it has to stay invariant under both simultaneous shifts, we must have 
\begin{equation}
Q_R(\delta,\theta)=Q_R(\delta/\lambda +\theta/2\pi).
\label{charge_dependence}
\end{equation}

\begin{figure}
\includegraphics[width=85mm]{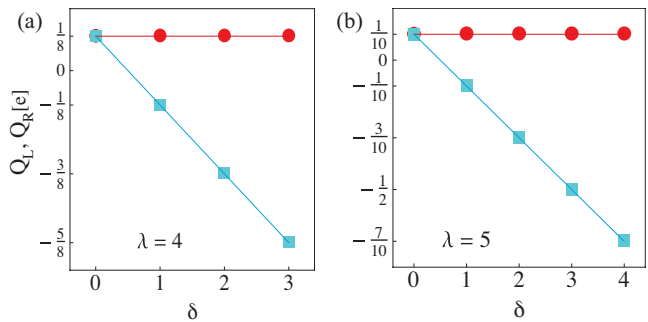}
\caption{(color online) The left $Q_L$ (red) and right $Q_R$ (blue) fractional boundary charges plotted as function of $\delta =(N+1)\ {\rm mod}\, \lambda$, which describes the extension of the array at the right end. The values  chosen in the numerics are:
 $\theta=0$,  (a) $\lambda = 4$, (b) $\lambda = 5$, and other parameters as in Fig.~\ref{fig:charge-theta}.
}
 \label{fig:charge-delta}
\end{figure}

\begin{figure*}
\includegraphics[width=170mm]{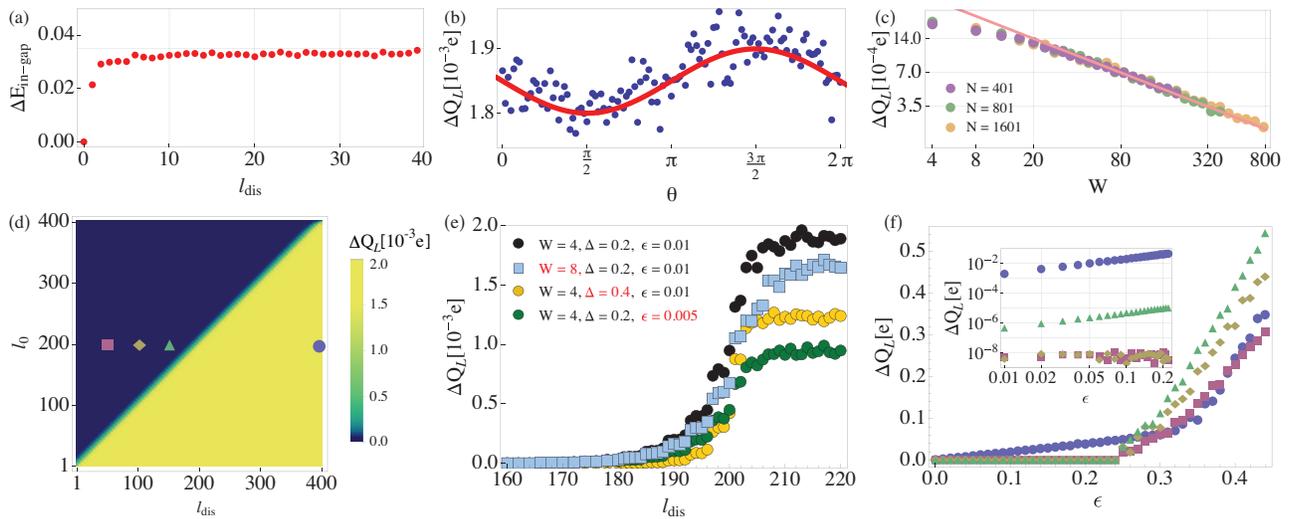}
\caption{(color online) Disorder induced fluctuations of (a) in-gap state energy for $\theta = \pi/2$, 
and (b)-(f) of the left boundary charge $Q_L$. 
Unless stated otherwise the parameters are $N=401$, $\lambda=4$, $\Delta/t=0.2$, $\epsilon/t=0.01$, $l_\textrm{dis}=N$, $W=\lambda$, $\theta=3\pi/2$, with each data point the standard deviation  $\Delta Q_L$ from 1000 random disorder configurations. 
(b) The red line is a fitted function $b-a \sin\theta$ with $b/a\approx 0.03$.
(c) Log-log plot with the line showing a $1/\surd{W}$ dependence.  
(e) Close-up on the crossover from stable to fluctuating charge, for parameters given at symbols. The black data correspond to a horizontal cut on panel (d) for $l_0=200$.
(f) Different symbols correspond to different $l_{dis}$ with the values denoted in panel (d). The inset is a log-log plot of the same numerical data. See text for explanations.}
 \label{fig:disorder}
\end{figure*}

From Eqs.~(\ref{delta_charge}) and (\ref{charge_dependence})  we can conclude that the functional dependence of $Q_R$ on $\delta$ and $\theta$ must have the form $Q_R = a_R -e (\delta/\lambda + \theta/2\pi)$, with $a_R$ 
determined below. We note that  Eq.~(\ref{charge_dependence}) is strictly speaking derived only for discrete values of $\theta$ such that $\theta$ changes in integer multiples of $ 2\pi \delta/\lambda$. This in turn implies that $a_R$ can still be a periodic function of $\theta$ with period $2\pi/\lambda$.
However, the deviations of $a_R$ from being constant become negligible in the continuum limit $\lambda\to\infty$, and already for $\lambda=4$ are very small, as shown by the numerical results plotted in Figs. \ref{fig:charge-theta} and ~\ref{fig:nonlinearity}. We therefore treat $a_R$ as a constant in what follows.

Next, let us determine the charge $Q_L$ on the left end. Again, as a local quantity $Q_L$ should not depend on $\delta$ and furthermore its dependence on the phase $\theta$ should 
be such as to cancel the $\theta$ term in
$Q_R$, since, obviously, the total charge $Q_R+Q_L$ must be invariant under changing $\theta$ modulo $e$. The jump by $e$ occurs when the in-gap state crosses the chemical potential
as function of $\theta$, but such jumps are irrelevant for the fractional part of the boundary charge.
To conclude, we arrive at $Q_L=a_L + e \theta/2\pi$.

To determine the constants $a_{L/R}$, we first use  Eq.~\eqref{eq:moae} which gives $a_L+a_R-e/\lambda=Q_{tot}-e(N+1-\delta)/\lambda$. For any $N$ the right hand side is an integer resulting in $a_L+a_R=e/\lambda$ (mod $e$).  A further condition on the constants can be  obtained by considering a symmetric configuration where $Q_R = Q_L$, which corresponds to $\theta = -\pi\delta/\lambda$, resulting in $a_R-a_L=0$. Eventually we obtain
\begin{align}
Q_L/e &=  {1 \over 2 \lambda} + {\theta \over 2\pi},
\label{eq:FCL}\\
Q_R/e &=  {1 \over 2 \lambda } - {\theta \over 2\pi} -{\delta \over \lambda}, 
\label{eq:FC}
\end{align}
which determine the fractional part of the boundary charges modulo $e$. 
In Fig.~\ref{fig:charge-theta}(a) the jumps in $Q_L$ and $Q_R$ by $e$ occur both at $\theta=\pi/2$, due to the particular choice of the chemical potential being at the degeneracy point of the left and right
in-gap states. Finally, the jump discontinuity ensures the $2\pi$ periodicity in $\theta$.

It is also interesting to consider  the continuum limit of the array and the associated charges. In this limit we need to retain only the leading order in the small parameter $a/\lambda_F=1/\lambda \ll 1$.
Applied to Eqs.~(\ref{eq:FCL},\ref{eq:FC}) this means that the boundary charges become in leading order $Q_L\approx {e\theta / 2\pi}$ and $Q_R \approx  - e{\theta / 2\pi}$ (modulo $e$). Note that this shows the characteristic linear dependence of the boundary charge on $\theta$ (now valid for any continuous $\theta$ value), which was found first~\cite{gw81} in the context of fractionally charged fermions in
the Jackiw-Rebbi model.~\cite{jr76} The latter model is related to our model in the continuum limit~\cite{Gangadharaiah2012}  at the chiral symmetry point corresponding to $\theta=\pi/2$ with an in-gap state of zero-energy and  fractional boundary charge  $e/4$ (modulo $e$). 
\footnote{We note that in the Jacki-Rebbi model the fractional charge bound to a kink in the gap is typically defined relative to the situation  with the same chiral symmetry but without in-gap bound state,~\cite{sharpness} which is the case here for $\theta=3\pi/2$ (corresponding to a repulsive potential value at the ends of the array). Thus, this relative fractional charge is given, e.g. at the right end, by $Q_R(\theta=\pi/2)-Q_R(\theta=3\pi/2)=e/2$ (modulo $e$).}

In addition to the tuning of the phase $\theta$, the boundary charges can be changed in discrete steps by varying the system size $N$ (and thus $\delta$). We note that in a spinful system, where all charges are doubled, at $\theta=0$ all rational fractions $Q_R=e(0,1,2,\dots,\lambda-1)/\lambda$ can be obtained for an odd integer $\lambda$, while only half of the rational fractions are available for an even integer $\lambda$.

\mysection{Fractional charge stability}
We now investigate the influence of disorder. To this end, we add to Eq.~\eqref{eq:H}
a term $\sum_{i=1}^{l_{\textrm{dis}}} \varepsilon_i c^\dag_i c_i$,
with random energies $\varepsilon_i \in \langle -\epsilon, \epsilon \rangle$, representing an uncorrelated on-site disorder of strength $\epsilon\geq 0$, extending from the left end of the array up to $l_\textrm{dis}$ dots. First, Fig.~\ref{fig:disorder}(a) shows the resulting fluctuations  (defined as the standard deviation) of the energy of an in-gap state for small disorder $\epsilon\ll |t|, |\Delta|$. Upon increasing $l_\textrm{dis}$ the fluctuations initially grow, saturating beyond $l_{dis}\approx 10$ where the disorder covers the whole in-gap state wave function, $l_{dis} \gtrsim \xi_n$, as numerically confirmed for $\xi_n(\theta=\pi/2)=10$ and shown in Fig.~\ref{fig:charge-theta}(b). 

As expected,  the fractional boundary charge $Q_L$ also fluctuates due to disorder, characterized by the root mean square value $\Delta Q_L$, for an example obtained from averaging over $1000$ random disorder configurations see  Fig.~\ref{fig:disorder}(b).
However, in contrast to  the in-gap state,  $\Delta Q_L$ depends only weakly on  $\theta$, which is the first indication that this fluctuation is of different nature. 
Figure \ref{fig:disorder}(c) shows that increasing the smoothness of the  profile function of $\hat{Q}_f$ (which can influence only the bulk of the array beyond $l_0=200$) suppresses the fluctuations, even though the localization of the boundary charge is only over few  sites $\xi_Q\approx 3$ [see Fig.~\ref{fig:charge-theta}(b)]. This suggests that the fluctuations of the boundary charges have strong contribution from density fluctuations in the bulk. In analogy to the operator $\hat{Q}_f$ sharpness (see App.~\ref{app:check}), one can define the intrinsic fluctuations of the boundary charge by subtracting the bulk contribution. Figure \ref{fig:disorder}(d) shows that these intrinsic fluctuations are exponentially small: the boundary charges are immune to disorder which reaches to any finite distance from the boundary, as long as it does not reach the support of the charge operator, $l_\textrm{dis}\leq l_0$. Figure \ref{fig:disorder}(e) zooms in on the crossover region $l_\textrm{dis}\approx l_0$ and shows the influence of various parameters: the fluctuations decrease if the gap $\Delta$ is increased, the disorder strength is decreased, or the profile smoothness is increased. Finally, Fig.~\ref{fig:disorder}(f) shows that the effect is not restricted to small disorder; the boundary charges are  stable (in the above sense) up to the disorder strengths $\epsilon$ of the order  $\Delta$. 

Equation \eqref{eq:FC} helps to understand how the fractional charge can be stable against strong disorder. Let us consider again a single-site potential fluctuation, this time inside the array. If it is very strong, it effectively removes the site leaving it either empty or occupied (depending on the potential sign) and cuts the array into two separate parts. This creates two new edges, where boundary charges will be induced. One quickly notices that Eqs.~\eqref{eq:FCL} and \eqref{eq:FC}  give the sum of these charges being $e/\lambda$, which exactly compensates the amount removed from  the bulk charge corresponding to a single site, $1\times \overline{\rho}=e/\lambda$. Physically, this reflects the stability of a band insulator where charges can not be displaced by a local potential to large distances. 
Importantly for practical realization, this also implies that the boundary charges are stable against disorder at the array ends, where typically the disorder might be stronger than inside the array. 

We finally note that these arguments apply for the fractional parts of the boundary charges. For the integer part to be stable, one has to make sure that the in-gap bound state is sufficiently far away from the chemical potential, otherwise disorder might push the bound state above  (below) the chemical potential and the bound state might get unfilled (or filled), which results in strong fluctuations (this is true even if  the total charge is fixed in a closed system). This is the case for $\theta$ close to a discontinuity of the boundary charge. Thus, also the total boundary charge is stable away from such discontinuities.

\section{Minimal array size}

\begin{figure}
\includegraphics[width=85mm]{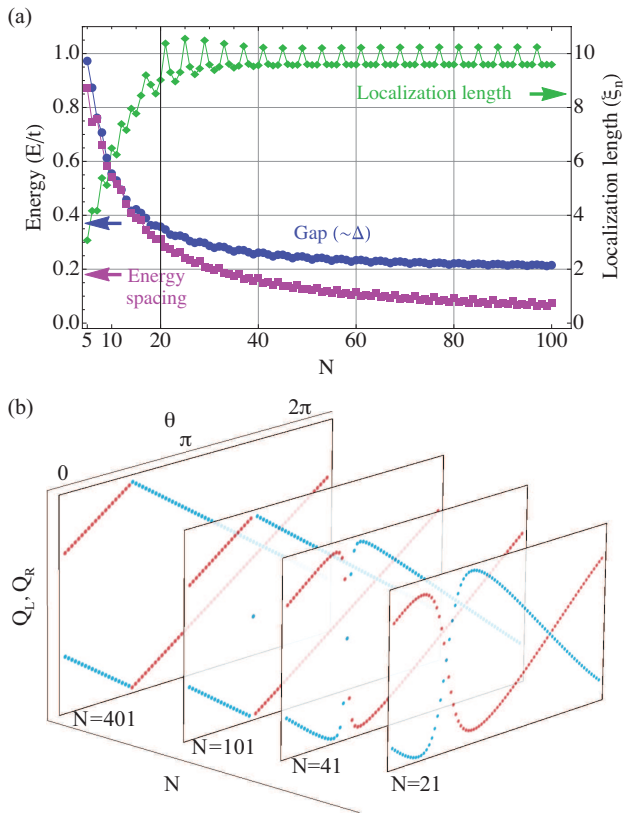}
\caption{(color online)
(a) Comparison of the band gap (blue curve; defined as the largest difference of two consecutive terms in an ordered energy spectrum) and the finite size quantization energy (purple curve; defined as the second largest difference), as functions of $N$ for $\theta = 2\pi-\delta\pi/\lambda$, which is a phase differing by $\pi$ from the phase at which the in-gap states cross in the lowest gap. The green curve shows the in-gap state localization length ($\xi_n$, defined as in Fig.~\ref{fig:spectrum}) 
at $\theta$ chosen such that the in-gap state energy is approximately in the middle of the gap. 
(b) Evolution of the fractional boundary charges $Q_{L,R}$ as function of $\theta$ for the  system of the size $N$, and with the other parameters being the same as in Fig.~\ref{fig:charge-theta}(a). 
} 
\label{fig:short}
\end{figure}

Let us comment on the experimental requirements on the fractional charge implementation.
Even though techniques of fabrication, tuning, and control of gated QDs are rapidly progressing~\cite{shinkai2009:PRL,medford2013:NN, delbecq2014:APL, takakura2014:APL, baart2015:NN,braakman2013:NN,amaha2012:PRB,delbecq2016:PRL,weperen2011:PRL}
, it is still very demanding to build long arrays. It is therefore of high practical importance to estimate the minimal required size of an array where fractional boundary charges and in-gap states could be established and probed. Figure \ref{fig:charge-theta}(b) suggests that an array of the order of ten QDs is sufficient, as the localization length of the boundary charges are very small (for the chosen value of the gap being an appreciable fraction of the tunneling energy). However, we note that in such short arrays there are additional complications. First, to place the chemical potential correctly, the position of the gap needs to be identified. In short arrays, this is not straightforward, as the states in the band have finite energy separations, due to the finite size energy quantization. Similarly, the in-gap states can be identified by their short localization length only if the latter can be clearly distinguished from the localization lengths of the states in the band, of the order of the system size. To demonstrate these finite size effects, in Fig.~\ref{fig:short}(a) we show how much the gap and the in-gap state localization lengths differ from the finite quantization energy and the system size, respectively. In addition, in Fig.~\ref{fig:short}(b) we show how the hybridization of the in-gap states results in a deviation  of the boundary charges from the linear behavior in an array with well separated ends. From all this  we conclude that an array with several tens of sites (QDs) is necessary, presumably $N=20$ as a minimum.

\mysection{Conclusions}

We have studied arrays of coupled QDs under periodically modulated onsite potentials. We found that fractional charges can be realized at the boundaries of the array, with values tunable by the phase of the on-site potential and the system size $N$. Our main results are that these fractional boundary  charges are independent of the in-gap bound states, their values described by Eqs.~\eqref{eq:FCL} and~\eqref{eq:FC}, and that they are stable against static on-site disorder. This suggests that the observation of  fractional boundary charges in arrays of QDs (or similar periodic structures) should be within experimental reach. In practice, a  single electron
transistor (SET) as a sensitive charge detector~\cite{yoo97,martin04,shach2015:PRB} may be used to perform measurements of fractional charges.

\acknowledgments 
We would like to thank G. Allison, S. Amaha, M. Delbecq, T. Nakajima, T. Otsuka, K. Takeda, S. Tarucha, L. Trifunovic, and J. Yoneda for helpful discussions. This work was supported by the Swiss NSF, NCCR QSIT, and by JSPS KAKENHI Grant Number 16H02204, 16K05411.

\appendix

\section{Fractional charges are well defined}

\label{app:check}

Here we demonstrate that the charge definition, Eq.~\eqref{eq:charge}, gives well defined fractional boundary charges. As discussed in the main text, there are two requirements to be satisfied: the charge value should be independent of the details of the profile function $f_i$ and the standard deviation of the operator $\hat{Q}_f$ in the ground state should vanish in the limit $W\to \infty$. We show now that these requirements are fulfilled.

\subsection*{Stability of the bulk density}

\begin{figure}
\includegraphics[width=85mm]{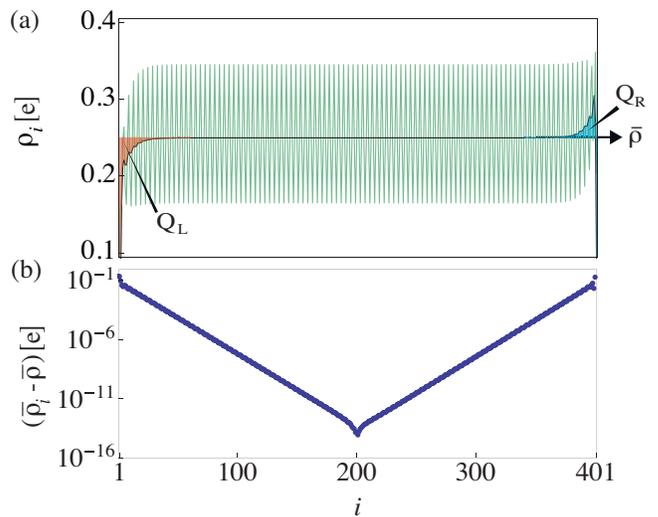}
\caption{(color online) 
(a) The ground state charge density $\rho_i$ (green) as a function of the dot index $i$ (position within the array). 
The density averaged over $\lambda$ dots, $\overline{\rho}_i$, converges to $\overline{\rho}$ (black horizontal line). The difference of $\overline{\rho}_i - \overline{\rho}$ plotted as red on the left and blue on the right, gives the boundary charges $Q_L$ and $Q_R$.
(b) The difference $\overline{\rho}_i-\overline{\rho}$ plotted in a log scale.
} 
\label{fig:density convergence}
\end{figure}

To show that the first requirement is met, we plot the ground state expectation value of the actual electronic density  $\rho_i=\langle c_i^\dagger c_i \rangle$, as the green line in Fig.~\ref{fig:density convergence}(a). This quantity displays oscillations with period $\lambda$, which are removed by averaging $\rho_i$ over the unit cell (meaning over $\lambda$ dots). In the bulk, the averaged value $\overline{\rho}_i$ is  equal to the constant $\overline{\rho}=e/\lambda$ (the black horizontal line) up to the numerical precision of our code (for large arrays; not shown), while the difference $\overline{\rho}_i-\overline{\rho}$ vanishes exponentially upon moving away from the boundary, as seen in Fig.~\ref{fig:density convergence}(b). 

The averaging of the local charge density can be effectively performed by averaging over the profile function $f$. For example, consider the left boundary charge defined using the locally averaged density $\overline{\rho}_i=(1/\lambda)\sum_{j=i}^{i+\lambda-1} \rho_j$, and an abrupt profile, $f_i^\prime=1$ for $i < l_0$ and 0 otherwise. Since both the averaging and the weighted summation are linear operations, they can be rewritten as a a single linear operation with a redefined profile,
\begin{equation}
\sum_i \overline{\rho}_i f_i^\prime = \sum_i \rho_i f_i,
\label{eq:averaging}
\end{equation}
with $f_i$ given in Eq.~\eqref{eq:profile} with $W=\lambda$. Similarly, starting with a profile $f_i^\prime$ with a linear drop over $n$ sites and an averaged local density $\overline{\rho}_i$, is equivalent to taking a non-averaged density $\rho_i$ and a profile drop $f_i$ with $W=n \lambda$.  
Having this in mind, in the main text we skipped introducing  the intermediate quantity $\overline{\rho}_i$ in defining the boundary charges at the expense of restricting the values of $W$ to integer multiples of $\lambda$. Evidently, this restriction becomes irrelevant in the limit $W\to \infty$, where the fractional boundary charges become sharp observables. 

\begin{figure}
\includegraphics[width=85mm]{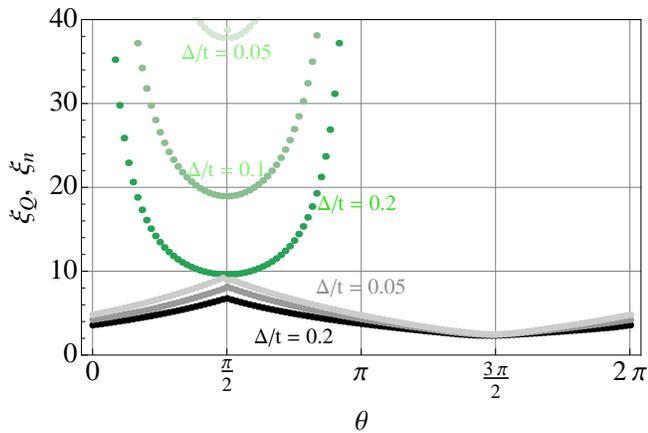}
\caption{(color online) Localization lengths of the boundary charge ($\xi_Q$ black) and the in-gap state ($\xi_n$; green) for three different values of the  potential amplitude $\Delta$ with values labeled and encoded by the curves hue. Other parameters are the same as in Fig.~\ref{fig:charge-theta}.} 
\label{fig:localization}
\end{figure}

We used the averaged charge density to characterize the localization length of the boundary charge by defining
\begin{equation}
\xi_{Q_f}^{-1} = \sum_{i=1}^N  |f_i \delta \overline{\rho}_i/e|^2 / (\sum_{i=1}^N |f_i \delta \overline{\rho}_i/e| )^2,
\label{eq:IPR}
\end{equation}
with the result plotted in Fig.~\ref{fig:charge-theta}(b) as an average for the left and right boundary charge $\xi_Q=(\xi_{Q_L}+\xi_{Q_R})/2$.\footnote{The averaging according to Eq.~\eqref{eq:averaging} resulting in prescription $\overline{\rho}_i=(\rho_{i-1}+\rho_i+\rho_{i+1}+\rho_{i+2})/4$ inside the array is terminated at the ends of the array by $\overline{\rho}_1=(\rho_1+\rho_2)/2$, and $\overline{\rho}_2=(\rho_1+\rho_2+\rho_3)/3$. Different choices lead to slightly different participation ratios with the same overall behavior as the one plotted in Fig.~\ref{fig:localization}.} For illustration we show in Fig.~\ref{fig:localization} how the localization lengths change with the potential amplitude $\Delta$ (equal to the gap). Interestingly, even though both the in-gap state and the boundary charge becomes less localized, as expected, the effect is much less pronounced for the latter quantity and does not scale inversely with the gap size, $\xi \propto 1/\Delta$, a relation which would hold for an in-gap bound state. Thus, again, this shows that,  in general, the fractional boundary charges are not directly related to in-gap bound states but instead come from all the filled states in the Fermi sea getting deformed at the boundary due to the vanishing boundary condition.

We checked that the fractional charges in Fig.~\ref{fig:charge-theta} are reproduced (using the formula on the left hand side of Eq.~\eqref{eq:averaging}; not shown) using an alternative, Gaussian, profile function,
\begin{align}
 f^{\prime \prime L}_i = \left\{ \begin{array}{ll}
         1, & \mbox{if $i<l_0$},\\
        \exp\left( -(i-l_0)^2/W^2 \right) , & \mbox{if $i\geq l_0 $}, \end{array} \right.
        \label{eq:profile2}
\end{align} 
where the parameters $l_0$ and $W$ have analogous meanings to those in Eq.~\eqref{eq:profile}. The independence of the boundary charges on the profile function follows directly from the fact that the difference $\delta\overline{\rho}_i=\overline{\rho}_i-\overline{\rho}$ is exponentially small, as shown in Fig.~\ref{fig:density convergence}. The independence is, however, less straightforward concerning the operator sharpness, which we discuss next.

\subsection*{Quantum fluctuations of fractional boundary charges}

\begin{figure}
\includegraphics[width=85mm]{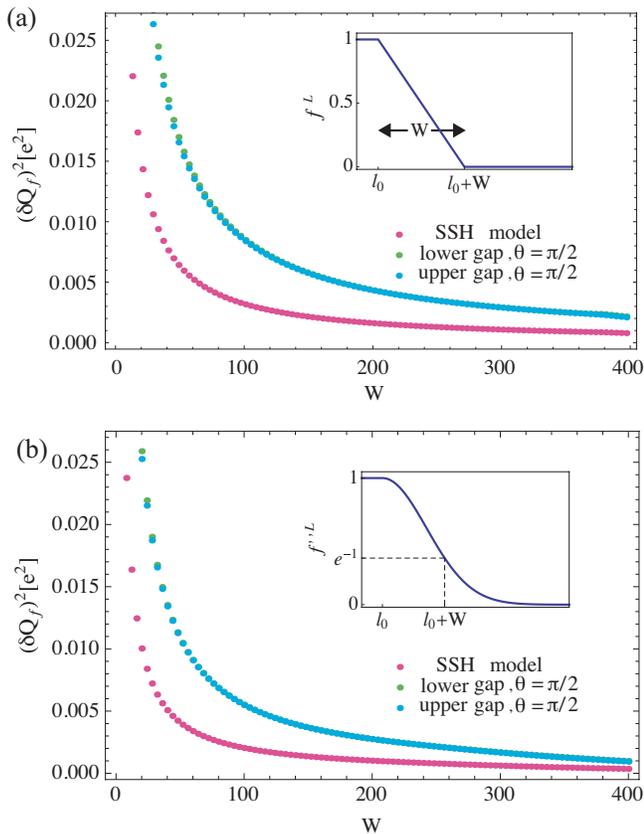}
\caption{(color online) Quantum fluctuations of left boundary charge q$\delta Q_f$ calculated for parameters given in Fig.~\ref{fig:spectrum}, $\theta=\pi/2$ and for (a) the piecewise linear profile function given in Eq.~\eqref{eq:profile}, and (b) a Gaussian function given in Eq.~\eqref{eq:profile2}, both with $l_0=20$. The chemical potential lies in the upper and lower band gaps as denoted at symbols. The fluctuations of the right boundary charge look very similar (not shown). All the plotted functions decay exponentially for large $W$, as we checked on a log plot (not shown). For comparison, we include also the results obtained using the SSH model,~\cite{ssh79} defined by 
$H = \sum_{j=1}^{N-1} [t+\Delta \cos (2\pi j/\lambda + \theta)][c^\dag_j c_{j+1} + c^\dag_{j+1}c_j]$ with $\lambda=2$, where the data are plotted for $\theta=0$, $\Delta/t = 0.2$, and $N = 400$.} 
\label{fig:charge-fluctuation}
\end{figure}

The simplest way how a non-integer mean charge can result is an average of several integer values. For example, a mean charge 1/2 can arise as an average of states with charges 0 and 1 with equal population probabilities. Measuring repeatedly, in this case one would obtain results such as, e.g., \{0, 0, 1, 0, 1, 1, 0, $\ldots$\}. This is different from a charge whose measurement results are \{1/2, 1/2, 1/2, 1/2, $\ldots$\}. Having the same mean, the two objects are distinguished by the charge operator standard deviation, also referred to as quantum fluctuations. A sharp fractional charge has, by definition, standard deviation $\delta Q_f$ much smaller than its non-integer ground state expectation value $Q_f=\langle \hat{Q}_f \rangle$. The former is defined by
\begin{equation}
(\delta Q_f)^2 = \langle (\hat{Q}_f - Q_f)^2 \rangle,
\end{equation}
which we can write at zero temperature as 
\begin{equation}
 (\delta Q_f)^2  =e^2 \sum_{n \in \textrm{occ.}} \sum_{m \notin \textrm{occ.}} \Bigl|\sum_{i=1}^N f_i\psi_{n,i}^*   \psi_{m,i} \Bigr|^2,
\end{equation}
with occ. standing for occupied states. 

To characterize the sharpness of the boundary charges in our model, we plot $\delta Q_f$ as functions of the boundary smoothness $W$ in Fig.~\ref{fig:charge-fluctuation}. The plots show that the standard deviation $\delta Q_f$ decays upon increasing $W$, that the decay is similar to that of a 1/2 boundary charge in the SSH model,~\cite{ssh79} and that the decay is not conditioned on a specific functional form of $f_i$. The last fact is shown by Fig.~\ref{fig:charge-fluctuation}(b) where a Gaussian profile given in Eq.~\eqref{eq:profile2} was adopted. The decay of quantum fluctuations in the limit $W\to \infty$ shows that the fractional boundary charges correspond to sharp quantum observables, rather than to an average of several integer values.

\subsection*{Linearity of the boundary charges}

\begin{figure}[h!]
\includegraphics[width=85mm]{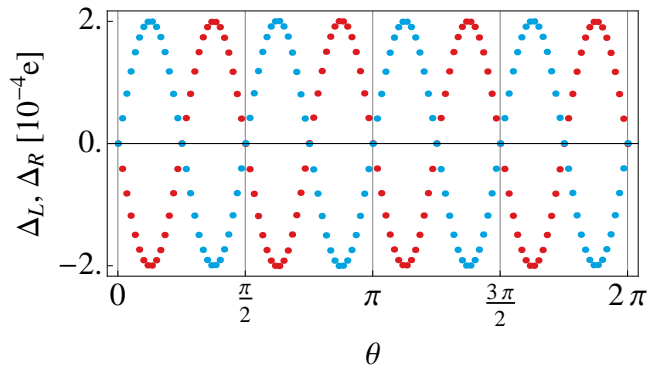}
\caption{(color online) The residuals of the boundary charges obtained from numerics as plotted in Fig.~\ref{fig:charge-theta} upon subtracting the analytical result given in Eq.~\eqref{eq:FC}, $\Delta_L=Q_L^{num}-Q_L$ mod $e$, and analogously for $\Delta_R$.} 
\label{fig:nonlinearity}
\end{figure}

In Fig.~\ref{fig:nonlinearity} we plot the oscillations of the quantities $a_L$ and $a_R$ upon changing the modulation phase $\theta$. These oscillations disappear in the continuum limit ($\lambda \to \infty$), however,  even for $\lambda=4$ they are already very small. This shows that considering the boundary charges linear in $\theta$ (for  any real value of $\theta$ mod $2\pi$) is an excellent approximation even for small $\lambda$.

\label{app:nonlinearity}

\begin{thebibliography}{99}
\bibitem{fnt1}
This means that they correspond to a sharp quantum observable, see Refs.~\onlinecite{sharpness}.
\end{thebibliography}
\end{document}